\documentclass[11pt]{article}
\newcommand{\Czero}[1]{\langle C(#1) \rangle_0} 
\newcommand{\Cmax}{C_{\rm max}}
\newcommand{\goesto}{\rightarrow}
\newcommand{\definedas}{\equiv}
\newcommand{\aboutequal}{\simeq}

\newcommand{\integral}{\int}
\newcommand{\goesas}{\sim}
\newcommand{\ie}{{\it i.e.}}
\newcommand{\eg}{{\it eg.}}
\newcommand{\cf}{{\it cf. }}
\newcommand{\newterm}[1]{{\sl #1}}

\usepackage{epsfig}
\begin{document}
%
%
\renewcommand{\thefootnote}{\fnsymbol{footnote}}
\title{{\Large Universal Ratios of Characteristic Lengths in Semidilute
Polymer Solutions}}
\author{Jung-Ren Huang\footnote{email: jhuang2@midway.uchicago.edu}
                  and T. A. Witten\\
{\em \small James Franck Institute and Department of Physics,
   University of Chicago} \\
{\em \small 5640 S. Ellis Avenue, Chicago, Illinois 60637}
 }
\date{May 10, 2002}
\maketitle
\begin{abstract}
We use experimental  and simulation data from the literature to infer five
characteristic lengths, denoted $\xi_s$, $\xi_f$, $\xi_\Pi$, $\xi_\phi$,
and $\xi_D$ of a semidilute polymer solution.  The first two of these are
defined in terms of  scattering from the solution, the third is  defined in terms of 
osmotic pressure, the fourth by the spatial monomer concentration profile,
and the last by co-operative diffusion.  In a given solution the ratios of
any of these five lengths are expected to be universal constants.  Knowing
these constants thus allows one to use one measured property of a solution
as a means of inferring others. We calculate these ratios and estimate
their uncertainties for solutions in theta as well as good-solvent
conditions.  The analysis is strengthened by use of scattering properties
of isolated polymers inferred from computer simulations.
\end{abstract}
%
\section{Introduction \label{intro}}
In the 1970's, it was recognized that polymer solutions are a form of critical 
phenomenon\cite{DeGennes.Physics.letters.1971}.
In the intervening years, a network of powerful consequences of this
recognition have been verified.  Many measured properties vary with
concentration and with molecular weight according to power laws whose
exponents are known only approximately, but which are shown to be
universal---\ie unchanged under continuous changes in the system properties.
Moreover, the coefficients in these power laws are shown to be
inter-related by universal ratios.
Some of these ratios predict universal properties of dilute solutions.  In
1987 Davidson et al expressed some of this universality in an elegant way
by determining characteristic lengths\cite{Dav}.
  Several experimental measurements
were made on a number of solutions, and each was reduced to a length.
Thus, \eg light scattering measurements were used to determine the radius
of gyration.  Several different lengths were inferred from different
measurements on a given solution.  When these lengths were compared, their
ratios were found to be independent of the solution studied, thus
confirming the anticipated universality.  In the process the experiments
obtained well-determined values for the universal ratios that remain as an
important part of our knowledge of polymers.

Our aim in this paper is to obtain analogous information for the semidilute
regime.  Semidilute solutions are commonly characterized by a length derived
from small-angle scattering called the correlation length
$\xi_s$\cite{deG1}.  Other common characteristics of these solutions can
also be reduced to a length.  For instance, the osmotic pressure $\Pi$ may be expressed
in terms of a length $\xi_\Pi$ defined by $\Pi \definedas k_B T /
\xi_\Pi^{\,3}$, where $k_B$ is Boltzmann's constant and $T$ the absolute
temperature. Likewise, the co-operative diffusion coefficient $D_c$ can be 
used to
define a hydrodynamic length $\xi_D$ analogous to Stokes' Law: $D_c \definedas
k_BT/(6\pi \eta_s \xi_D)$, where $\eta_s$ is the solvent viscosity.  Like
 Davidson's dilute lengths\cite{Dav}, the ratios of these $\xi$'s are expected to be
universal in the semidilute limit. The semidilute limit means the limit in
which
the volume fraction of polymer is at once much larger than the overlap
volume fraction, and much smaller than unity.  Naturally this limit
requires polymers of sufficiently high molecular weight.

In this paper we
determine these semidilute length ratios and several others using data from the
experimental literature. We expect that knowledge of these ratios should be
useful for those who study these solutions.  Experimentally, one may use the
ratios to predict one experimental quantity such as the co-operative
diffusivity
from another, such as the scattering correlation length.  Conceptually, these
ratios give one a clearer picture of the interior structure of the
solution.  We
emphasize that these predictions are not scaling relations with undetermined
numerical prefactors.  They are quantitative predictions with stringent
uncertainly limits.   

The universal properties of semidilute solutions have been
much explored and tested over the past two decades.  The purpose of our work is
not to extend the scope of these tests.  Instead, we aim to distill known
semidilute results into a form that is as useful and simple as possible for one
studying a particular semidilute solution.  Thus we define our lengths in a way
that does not require a knowledge of the dilute properties of the polymer
and solvent in question.
Another virtue of our approach is accuracy. Most universal ratios  for
semidilute quantities reported in the literature require 
knowledge of dilute quantities with the same polymer 
and solvent, such as the radius of gyration. These dilute quantities are often sensitive
to polydispersity, while the semidilute quantities of interest are not.
Thus, recasting the universal information without reference to dilute quantities removes
an important source of uncertainty.

%
Before proceeding, we define explicitly the lengths we will discuss.
As noted above, the \newterm{osmotic} length $\xi_{\Pi}$ is related to the
osmotic pressure $\Pi$ of the polymer solution by
 \begin{equation}
\Pi=k_B T/\xi_{\Pi}^{\,3}, \label{Pixi}
\end{equation}
where $k_B$ is the Boltzmann constant and $T$ the temperature.
We also define a \newterm{diffusive} length $\xi_D$ from the
co-operative diffusion coefficient $D_{c}$\cite{deG1, Doi}:
\begin{equation}
  D_{c} \definedas \frac{k_B T}{6\pi \eta_s \xi_D},   \label{Dch}
\end{equation}
where $\eta_s$ is the viscosity of the solvent.
The \newterm{scattering} correlation length $\xi_s$ is inferred from the
static structure
factor $S(q)$ at wavevector $q$\cite{PiS}:
\begin{equation}
S(q)=S_0(1-\xi_s^2q^2+{\cal O}(q^4)),  \label{TEs}
\end{equation}
where $S_0$ is the extrapolation of $S(q)$ at $q=0$. 
We note that $\xi_{s}^{2}=\frac{1}{3}R_G^2$ in the dilute limit,
where $R_G$ is the radius of gyration of the polymer.
A related length $\xi_f$ may be inferred from the scattering in the
so-called \newterm{fractal} wavevector regime where $1/q$ is much
smaller than $\xi_s$ but much larger than a monomer.  In this regime $S(q)
\goesas q^{-1/\nu}$, where the Flory swelling exponent $\nu \aboutequal 0.588$ for
good-solvent cases\cite{Schafer}.         
From this fractal
law, we may define the length $\xi_f$ by
\begin{equation}
S(q)\goesto S_0 (q \xi_f)^{-1/\nu}
\label{xi.f.definition}
\end{equation}
where $q$ is in the fractal regime.
Here ${\bf a} \goesto {\bf b}$ means that ${\bf a}$ approaches ${\bf b}$ in
the asymptotic limit under discussion.

Closely related to $\xi_f$ is a length characterizing the local monomer
concentration profile.  We define $C$ to be the monomer concentration and
$\Czero{r}$ as the ensemble average of the
\newterm{local} concentration at distance $r$ from an arbitrary monomer. If
this $r$ is much larger than the monomer size $\{a\}$ and much smaller than $\xi_s$,
 then $\Czero{r} \goesas r^{1/\nu - 3}$\cite{PiS}. This behavior gives rise to the fractal
scattering of Eqn \ref{xi.f.definition}, as discussed below.  We define the
\newterm{concentration} length $\xi_\phi$ by
\begin{equation}
\Czero{r}\goesto C (r/\xi_\phi)^{1/\nu - 3}
\label{xi.phi.definition}
\end{equation}
where $r$ is in the fractal regime defined above.  That is, $\xi_\phi$
is the distance at which the extrapolated fractal concentration profile
meets the solution concentration $C$, as shown in Figure \ref{figphi}. 
\begin{center}
\begin{figure}
\unitlength 0.5mm
\begin{picture}(260,150)
\put(75,10)
{
  \begin{picture}(180,135)
    \put(0,-2){\vector(0,1){114}}
    \put(-8,0){\vector(1,0){160}}
    \multiput(0,15)(6,0){23}{\line(1,0){3}}
    \multiput(85,15)(0,-4){5}{\line(0,-1){2}}
    \multiput(20,80)(0,-4){21}{\line(0,-1){2}}
    \multiput(-2,102)(2,-2){50}{\circle*{0.5}}
    \put(120,15){\line(1,0){20}}
    \put(20,80){\line(1,-1){40}}
    \qbezier(60,40)(85,15)(125,15)
    \qbezier(7,88)(13,85)(20,79.8)
    \put(35,65){\line(1,0){10}}
    \put(45,65){\line(0,-1){10}}
\put(5,118){\makebox(0,0){$\log \Czero{r}$}}
\put(146,-5){\makebox(0,0){$\log r$}}
    \put(85,0){\line(0,-1){2}}
\put(86,-6){\makebox(0,0){$\xi_\phi$}}
    \put(20,0){\line(0,-1){2}}
\put(20,-6){\makebox(0,0){$\{a\}$}}
\put(-15,90){\makebox(0,0){$\{\Cmax\}$}}
    \put(0,91){\line(-1,0){2}}
\put(-7,14){\makebox(0,0){$C$}}
    \put(0,15){\line(-1,0){2}}
\put(48,68){\makebox(0,0){$1/\nu-3$}}
  \end{picture}
}
\end{picture}
\caption{\label{figphi} 
Schematic Diagram of $\Czero{r}$(solid line) and
 $\{\Cmax A^{3-1/\nu}\}/r^{3-1/\nu}$(dotted line).}
\end{figure}
\end{center}

We note here that another dynamic length $\xi_p$ can be defined as the radius of a 
circular pipe with the same solvent permeability as 
the polymer solution of interest\cite{Tomsbook}.
The solvent permeability $P$ is usually measured with
 sedimentation experiments\cite{Mij,Vid}.
The sedimentation coefficient, however,  is  rigorously related to 
  the co-operative diffusion coefficient $D_c$ and the osmotic 
pressure $\Pi$\cite{Nystrom,Ada}.
Hence, $\xi_p$ can be expressed in terms of  $\xi_D$ and $\xi_\Pi$\cite{Tomsbook}:
\begin{equation}
  P \equiv { \xi_p^2 \over 8} = {{3-1/\nu}\over {18\pi}}{\xi_\Pi^3\over \xi_D}.
  \label{Pxip}
\end{equation}

This paper is organized as follows:
In Section \ref{LSE} we express the static lengths, $\xi_s$, $\xi_\Pi$,
$\xi_f$ and $\xi_\phi$, in terms of the
basic quantities $\underline{\beta_2}$, $\underline{\beta_4}$, $\{A\}$ 
and $\underline{P_\infty}$. 
 In Section \ref{DataS} we describe the experiments
and simulations on which our results are  based.  In Section \ref{Results}
we report our values of the $\xi$ ratios for both good and theta solvents.
In Section \ref{Discussion} we comment on the limitations and implications
of these ratios.
To improve the readability of the paper, 
we indicate quantities that depend only on polymer and
solvent species but not on concentration or molecular weight 
by curly-bracketed symbols, such as $\{A\}$,
  and quantities that are universal by underlined symbols, such as 
$\underline{P_\infty}$. 
We treat as universal any ratio of two quantities a) that individually diverge in the semidilute
 limit defined above, and b) that have the same predicted scaling dependence on a parameter 
such as concentration, so that their ratio is predicted to be independent of the parameter.
%
%
%
\section{Relations to Structural Coefficients \label{LSE}}
%
In this section we define a set of structural coefficients that
characterize the concentration profile in a semidilute solution.  We found
it convenient to relate the static lengths, $\xi_s$, $\xi_\phi$, $\xi_f$ and $\xi_\Pi$
 to these coefficients.
\subsection{Local Concentration}
Above we defined the length $\xi_\phi$ from the local concentration
$\Czero{r}$.  We may express this $\Czero{r}$ in terms of the bulk
concentration $\{\Cmax\}$ for the polymer in question.  Here we exploit the
fact that $\Czero{r}$ is unaffected by the surrounding solution when $r$ is
small.  Thus in the fractal $r$ regime,
we may use Eqn \ref{xi.phi.definition} and define the \newterm{fractal length} $\{A\}$:
\begin{equation}
{\Czero{r} \over \{\Cmax\}} \goesto \left( \frac{\{A\}}{r} \right)^{3-1/\nu}, \: \mbox{for
 $\{a\} \ll r \ll \xi_{\phi}$,} \label{Czero}
\end{equation}
where $\{a\}$ is the monomer size.
This $\{A\}$ is useful because it is independent of concentration.  
Evidently,
\begin{equation}
\xi_\phi = \left(\frac{\{\Cmax A^{3-1/\nu}\}}{C}\right)^{\nu/(3\nu - 1)}.
\label{xi.phi.A.ratio}
\end{equation}
The coefficient $\{\Cmax A^{3 - 1/\nu}\}$
depends only on the polymer and solvent species and is
independent of concentration; 
The scattering structure function $S(q)$ in the fractal $q$ regime can also
be expressed in terms of the fractal length $\{A\}$.  Here we take the monomer
as our elementary scatterer.  For a scattering volume $V$, $S(q)$ may
be written as\cite{PiS}
\[
  S(q) = \frac{1}{ V C}\sum_{j,k=1}^{V C} \langle \exp\left[ i \vec{q}\cdot
\left( \vec{r}_j-\vec{r}_k\right) \right]\rangle
\]
where $\vec{r}_j$ is the position vector of the $j$th monomer and $\vec{q}$ the
wave vector transfer.
In our notation, this reduces to
\begin{equation}
 S(q) = \int_{V} d^{3}r \Czero{r} \exp (i\vec{q}\cdot\vec{r}),
\label{sofq.czero}
\end{equation}
The fractal regime of $\Czero{r}$ gives rise to a complementary behavior in
the fractal $q$ regime\cite{PiS}:
The power law regime of Eqn \ref{Czero} produces the scattering at
fractal wavevector mentioned in Eqn. \ref{xi.f.definition}.  By
combining Eqn \ref{Czero} and \ref{sofq.czero}, one finds
\begin{equation}
S(q) \goesto c_{_0} \frac{\{\Cmax A^{3 - 1/\nu} \}}{q^{1/\nu}}\:\: \mbox{ for
 $1/\xi_\phi \ll q \ll 1/\{a\} $,} 
\label{Siq0}
\end{equation}
where $c_{_0} =4\pi \sin(\frac{\pi}{2}(1/\nu-1)) \Gamma(1/\nu -1)$.  We note
that Eqn \ref{Siq0} still holds in the dilute limit, where $\xi_s
\aboutequal R_G$, so the fractal $q$ regime means $1/R_G \ll q \ll 1/\{a\}$.

 For a given polymer and solvent, the coefficient $\{\Cmax A^{3 - 1/\nu}\}$
 can also be related
to the dilute property of that polymer solution.  The chainlength $N$ of an isolated
 polymer is related to the radius of gyration 
$R_G$ by $N \goesto (R_G/\{\Omega\})^{1/\nu}$.
We follow the work of Rawiso et al and define a universal constant 
$\underline{P_\infty}$ for a dilute polymer solution\cite{Raw}:
\begin{equation}
\underline{P_\infty} \goesto S(q) /N  \cdot  (q R_G)^{1/\nu}\:\: \mbox{ for
 $1/R_G \ll q \ll 1/\{a\}$.} 
\label{Pinfty}
\end{equation}
$\{\Omega\}$ can then be written as  
\begin{equation}
\{\Omega\} = (\underline{P_\infty}/c_{_0})^\nu 
        \{\Cmax A^{3 - 1/\nu}\}^{-\nu}.
\label{Omega.of.A}
\end{equation}
We note that $\underline{P_\infty}$ is  just 
$\bar{P}_\infty$ defined in the paper of Rawiso et al\cite{Raw}.
Since $\{\Omega\}$ and $\{\Cmax A^{3 - 1/\nu}\}$ are both properties of a single
chain, we can determine $\underline{P_\infty}$ using
single-chain experiments or simulations.
%
\subsection{Moments of Local Concentration \label{betas}}
%
%
Beyond the fractal regime governed by the coefficient $\{A\}$ discussed above,
the local concentration $\Czero{r}$ departs from the power-law form and
becomes constant(Figure \ref{figphi}).  It is convenient to define the 
reduced moments $\underline{\beta_n}$
to characterize this part of the concentration profile:
\begin{equation}
\underline{\beta_n} \definedas {\integral_0^\infty dr r^n (\Czero{r} - C) \over 
\integral_{C_A > C} dr
r^n (C_A(r)-C)} ,\label{def.beta.n}
\end{equation}
where $C_A(r) \definedas \{\Cmax\} (\{A\}/r)^{3- 1/\nu}$, and
 we use the behavior in the fractal regime to normalize these moments.
Several of the $\xi$ ratios can be expressed completely in terms of these 
reduced moments. 
By Taylor-expanding Eqn \ref{sofq.czero} with respect to $q$,
we show in the appendix that
for semidilute polymer solutions(Eqn \ref{sphi}),
\begin{equation}
    \underline{\xi_s /\xi_\phi} \goesto
\left[\frac{\underline{\beta_4}}{10(1+2\nu)\underline{\beta_2}}\right]^{1/2}.
    \label{XIS}
\end{equation}
The moment $\underline{\beta_2}$ may be related to $\xi_\Pi$ by using 
the compressibility sum rule\cite{deG1}:
\begin{equation}
S_0=k_B T\frac{\partial C}{\partial \Pi}, \label{S0Pi}
\end{equation}
where $S_0$ is defined in Eqn \ref{TEs}.
This $S_0$ can then be expressed in terms of $\underline{\beta_2}$,
 thus yielding a relation between
 $\underline{\xi_\Pi / \xi_\phi}$ and $\underline{\beta_2}$(Eqn \ref{ratio.xipi.xiphi}): 
\begin{equation}
 \underline{\xi_\Pi/\xi_\phi} \goesto  \left( 4\pi\nu \underline{\beta_2} \right)^{1/3}.
\label{Piphi}
\end{equation}
Finally, by virtue of Eqn \ref{xi.f.definition}, 
\ref{Siq0} and \ref{S00},
we obtain
\begin{equation}
  \underline{\xi_f/\xi_\phi}\goesto \left[ \frac{4\pi}{3c_{_0}} (3\nu-1) 
\underline{\beta_2} \right]^\nu, \label{xif}
\end{equation}
where the numerical coefficient $c_{_0}$ is defined below Eqn \ref{Siq0}.
We hence conclude that $\xi_f$ gives no additional information if $\xi_\phi$ 
and $\xi_\Pi$ are known.
%
%
%
\section{Sources of Data\label{DataS}}
\begin{table}
\begin{center}
 \parbox{4.8in}{\caption{\label{tabdata}Sources of experimental and simulation data used in this work.
  }}
\begin{tabular}{|l|c|c|ccccccc|}
\hline
  Source$^*$ & Probe$^{**}$ & $M_w\times 10^{-6}$ & $\underline{P_\infty}$ & $\{\Omega\}$ & $\xi_\phi$ & 
   $\xi_s$ & $\xi_\Pi$ & $\xi_D$ & $\xi_f$ \\
\hline
\hline
  dm1\cite{Destree} & MC & $N\leq4096$ & $\surd$ & $\surd$&-&-&-&-&- \\ 
  dm2\cite{Muller} & MC & $N\leq2048$ & $\surd$ & $\surd$&-&-&-&-&- \\
\hline
  da\cite{Hay} & LS & 8-60 & - & $\surd$ & $\surd$ & - & - & -&- \\
  sa\cite{Ada,Ste,Picot} & LS & 0.4-21 & - & - & - & $\surd$ & - & -&- \\
  sa\cite{Ste} & LS & 1-21 & - & - & - & -  & $\surd$ & - &-\\
  sa\cite{Ada,Roo} & DLS, CGD & 0.4-21 & - & - & - & - & - & $\surd$&- \\
\hline
  db\cite{Raw} & NS & $0.05-1.3 $ & $\surd$ & $\surd$ & $\surd$ & - & - & -&- \\
  sb\cite{PiS,Farnoux} & NS & 0.5-1 & - & - & - & $\surd$ & - & -& $\surd$  \\
\hline
  dc\cite{Higo} & LS & 0.2-2 & - & $\surd$ & $\surd$ & - & -  & -&- \\
  sc\cite{Ham} & XS & 0.1-2 & - & - & - & $\surd$ & $\surd$  & -&- \\
  sc\cite{Han} & CGD & 0.2-3 & - & - & - & - & - & $\surd$ &-\\
\hline
\end{tabular}
\parbox{4.8in}{
\begin{description}
\item{$*$}: 
 ``d'' and ``s'' stand for ``dilute'' and ``semidilute'', respectively.
  m1:  a polyethylene chain model of $N$ C-C bonds.
  m2:  the bond-fluctuation lattice  model for a polymer of chainlength $N$.
  a: PS+Cyclohexane at 35$^\circ$C. 
  b: PS+CS$_2$ at 20$^\circ$C.
  c: PS+Toluene at 25$^\circ$C.
\item{$**$}: 
 MC: Monte Carlo Simulation.
 LS: Light Scattering.
 NS: Neutron Scattering. XS: X-ray Scattering. DLS: Dynamic LS.
 CGD: Classical Gradient Diffusion. 
\end{description}
}
\end{center}
\end{table}
%
%
Our results are based on the experimental data
 of solutions of polystyrene(PS) in the good solvents,
toluene and carbon disulfide(CS$_2$), and the solvent cyclohexane at its
theta temperature. 
For non-deuterated PS, the molecular weight $\{m_s\}$ of the 
monomer (CH$_2$CHC$_6$H$_5$) is 104 and the bulk density $\{\rho_{\rm max}\}$ is
 $1.05$g/cm$^3$\cite{handbook},
so that $\{\Cmax\}=6.08\times 10^{-3} \mbox{\AA$^{-3}$}$. 
In this paper, we assume the value of $\{\Cmax\}$ is invariant, 
regardless of whether the polymers are deuterated or not.
Thus,
we have ignored possible effects of tacticity of the molecules or of processing to 
obtain the neat (glassy) state of the polymer. Such effects introduce insignificant 
uncertainties compared to other uncertainties.
 Table \ref{tabdata} summarizes the sources of data  and the
quantities we derive from the associated sources. 
%
%
%
\subsection{The Values of $\underline{P_\infty}$\label{sofchi}}
The determination of $\xi_\phi$ requires the knowledge of  
$\{\Cmax A^{3-1/\nu}\}$, which can be obtained from the scattering data in 
the fractal $q$ regime(Eqn \ref{Siq0}).
However, data of this sort are rare.
Measurements of $\{\Omega\}$(Eqn \ref{Omega.of.A}) are much more accessible 
and provide the same information  if  the universal 
coefficient $\underline{P_\infty}$ is known.  
We may determine $\underline{P_\infty}$ using polymer simulations or theory. 
In  theta conditions $\underline{P_\infty}$ may be determined analytically by
treating the polymers as ideal random walks\cite{Cotton}.
 It is straightforward to show that $\underline{P_\infty}=2$ for theta cases\cite{Raw}.
Destr\'ee et al obtained $S(q)$ for a single polyethylene in 
theta and good-solvent conditions using Monte Carlo techniques\cite{Destree}.
From their results,
we determine $\underline{P_\infty}=2(\pm 4\%)$ for the theta cases where
the number of C-C bonds $N\leq 4096$,
and $\underline{P_\infty}=1.26(\pm 22\%)$ 
for the  good-solvent cases where $N\leq 1024$.
M\"uller et al reported $S(q)$ in their  Monte Carlo 
simulation of a single chain\cite{Muller}, which yields
 $\underline{P_\infty}=1.23(\pm 4\%)$.
 $\underline{P_\infty}=1.20(\pm 14\%)$ was also obtained using
the scattering data for dilute PS in the good solvent CS$_2$\cite{Raw}.
These values are consistent with the renormalization group estimate of
 $\underline{P_\infty}=1.29$ to first order in $\epsilon\equiv $(4 minus the dimension of space)\cite{Schafer,Wit}.
In the following sections,  $\underline{P_\infty}=2$ and $1.23(\pm 4\%)$ are
 adopted to calculate $\xi_\phi$ for theta and good-solvent cases, respectively.
%
%
\subsection{PS+Cyclohexane at 35$^\circ$C}
Although for many purposes, polymers in theta solvents may be considered 
as ideal random walks,
 theta polymers interact, and their semidilute solutions differ from a 
solution of noninteracting polymers. Thus not all of the universal
 ratios we seek for 
theta solutions can be found analytically. Here we make use of several studies of 
 PS in cyclohexane at 34.5$^\circ$C(or 35$^\circ$C), a well-known theta solvent.
 Hayward and Graessley reported 
$R_G \goesto 0.28(\pm4\%) M^{0.5}\:\mbox{\AA}$\cite{Hay},
which implies
\begin{equation}
\{\Omega\} = 2.86 (\pm4\%)\:\: \mbox{\AA}. \label{Omegacyc}
\end{equation} 
The behavior of  $\xi_s$  is given in Adam and Delsanti's paper\cite{Ada},
where they summarized the results of neutron and light scattering 
experiments\cite{Ste,Picot}: 
\begin{equation}
\xi_s = \frac{5.5}{\rho}(\pm5\%)\:\mbox{\AA}, \label{xiscyc}
\end{equation}
where $\rho$ is measured in g/cm$^3$.
They also reported
\begin{equation}
D_c = 1.25(\pm 8\%) \times 10^{-6}\rho\:\:\:
\mbox{cm$^2$/s},\:\:\mbox{for $\rho\leq 0.08$ g/cm$^3$}.
  \label{xidcyc}
\end{equation}
We note that the viscosity $\eta_s$ of cyclohexane at 35$^\circ$C is 
$0.762\times 10^{-2}$ poise\cite{Han}.  
Their result is consistent  with the classical gradient diffusion measurement done by 
Roots and Nystr\"om\cite{Roo}. 
Stepanek et al measured the osmotic compressibility using light scattering
 technique\cite{Ste}. They obtained
\begin{equation}
\frac{1}{\rho}\frac{\partial \Pi}{\partial \rho} = 2.93\times 10^7 \rho\:\:\:
         \mbox{for } 3< \rho/\rho^* < 11,
\label{MRTPicyc}
\end{equation}
where $\partial \Pi/\partial \rho$ is expressed in dyn$\cdot$cm/g and 
$\rho$ is in g/cm$^3$.
%
%
\subsection{Perdeuterated PS+CS$_2$ at 20$^\circ$C\label{PSCS2}}
CS$_2$ is known to be a good solvent for PS at 20$^\circ$C.
Rawiso et al measured 
the radius for dilute perdeuterated PS in CS$_2$ and reported 
$R_G \goesto 0.133(\pm 4\%)M_w^{0.588}$(above Eqn (46) of \cite{Raw}).
Since the molecular weight $\{m_s\}$ of a perdeuterated monomer is 112, 
according to the definition of $\{\Omega\}$ above Eqn \ref{Pinfty}, we infer
\begin{equation}
 \{\Omega\} = 2.13(\pm 4\%) \:\:\mbox{\AA}.   \label{lamPSCS2}
\end{equation}
Here $M_w$ is identified with $M$ without further concern about the polydispersity,
which is justified if the polydispersity is small.
 Rawiso et al also reported $S(q)$ in the fractal $q$ regime (above Eqn(46) of \cite{Raw}):
\[
  S(q)\cdot q^{1/\nu} \goesto 0.330(\pm 8\%)\:\: \mbox{\AA}^{-1/\nu}, 
\]
which in view of Eqn \ref{Siq0}, gives
\begin{equation}
\{\Cmax A^{3-1/\nu}\} = 2.27(\pm 8\%)\times 10^{-2} \:\: \mbox{\AA}^{-1/\nu}.
 \label{iqPSCS2}
\end{equation}
Farnoux et al measured $\xi_s$ for semidilute perdeuterated PS in CS$_2$ 
 also using neutron scattering \cite{Farnoux}:
$\xi_s=44\mbox{\AA} (\rho=0.025\mbox{ g/cm$^3$}, M_w=1.1\times 10^6)$,
 $29.8\mbox{\AA}  (\rho=0.04, M_w=5\times 10^5)$, $18.3\mbox{\AA}  (\rho=0.075, M_w=5\times 10^5)$ 
and $10.5\mbox{\AA}  (\rho=0.15, M_w=5\times 10^5)$. 
We note that since they used perdeuterated PS,
 the monomer concentration $C$ and the chainlength $N$ are related to 
$\rho$ and $M$ by $C=N_a \rho/112$ and $N = M / 112$, 
respectively, where $N_a$ is Avogadro's constant.
des Cloizeaux and Jannink reported in their book(below Eqn (15.4.37) of \cite{PiS}) that
\begin{equation}
\frac{S_0}{S(q)}\cdot (q\xi_s)^{-1/\nu}= 1.51(\pm1\%)\:\: 
\mbox{ for  $\:q\xi_s\geq 2.6$},  \label{Sxif}
\end{equation}
from which we can calculate $\xi_f$.
%
%
%
\subsection{PS+Toluene at 25$^\circ$C}
Higo et al summarized the results of several light scattering experiments 
and reported $R_G^2=1.38\times 10^{-2} M^{1.19}$\AA$^2$\cite{Higo}.
This implies
\begin{equation}
\{\Omega\} = 1.95(\pm 5\%)\:\:\mbox{\AA} \label{Omegatol}
\end{equation}
The small-angle X-ray scattering experiment done by Hamada et al gives
$\xi_s$ and  $S_0$\cite{Ham}:
\begin{equation}
\xi_s =  2.67 {\rho}^{-0.77}(\pm5\%)\:\mbox{\AA},
 \label{xisrhotol}
\end{equation}
and
\begin{equation}
 S_0 = 2.65 \rho^{-1.30}(\pm10\%), \label{Sxpirhotol}
\end{equation} 
where $\rho$ is in g/cm$^3$.
The co-operative diffusion constant is given in Schaefer and Han's 
review article\cite{Han}:
\begin{equation}
D_c = 93.0\cdot \left({C \over \{\Cmax\}}\right)^{0.75} (\pm5\%)\cdot 10^{-7}
\mbox{cm$^2$/s},\:\:\mbox{for $0.02 \leq C /\{\Cmax\}\leq 0.08$}.
  \label{dcc}
\end{equation}
We note that the viscosity $\eta_s$ of toluene at 25$^\circ$C is 
$0.552\times 10^{-2}$ poise\cite{Han}.
%
%
\section{Results\label{Results}}
We report in this section the ratios of  $\xi_{s}$, $\xi_{\phi}$, 
$\xi_{\Pi}$, $\xi_f$, and $\xi_D$, as well as the reduced moments $\underline{\beta_2}$ 
and $\underline{\beta_4}$ 
for polymer solutions under theta and good-solvent conditions.
$\xi_f$ is rigorously related to a combination of $\xi_\phi$ and 
$\xi_\Pi$(Eqn \ref{xif}), and 
$\xi_p$ can be determined using Eqn \ref{Pxip}.
For PS+Cyclohexane and PS+Toluene solutions,
 we are able to determine all $\xi$'s except $\xi_f$.
But for PS+CS$_2$ solution,
 we obtain $\xi_{\phi}$, $\xi_s$ and
$\xi_f$, but find  no  data in the literature helpful in determining $\xi_{\Pi}$ and
$\xi_D$.
The results are also summarized in Table \ref{tabratio} for clarity's sake.
The quoted uncertainties are those claimed in the original references. 
 When our quantity involves more than one number from these references, 
we have propagated the quoted uncertainties presuming that these are independent. 
\begin{table}
\begin{center}
 \parbox{4.8in}{\caption{\label{tabratio}Universal Ratios of the
Characteristic Lengths in Semidilute Polymer Solutions.
}} 
\hspace*{-1.1cm}
\begin{tabular}{|c|c|cccc|cc|}
\hline
System$^{*}$  & $\{A\}$(\AA) & $\underline{\xi_\phi / \xi_s}$ & $\underline{\xi_\Pi / \xi_s}$ & 
  $\underline{\xi_f / \xi_s}$  &  $\underline{\xi_D / \xi_s}$ 
     & $\underline{\beta_2}$ & $\underline{\beta_4}$ \\
\hline
 a & $3.21(\pm 7\%)$ & $0.61(\pm9\%)$ &
   $2.97(\pm5\%)$ & - & $4.29(\pm10\%)$ & $18(\pm22\%)$ & $968(\pm28\%)$ \\
b  &$ 2.79(\pm7\%)$ & $1.23(\pm8\%)$ & - & $1.27(\pm1\%)$ & - & $4.8(\pm14\%)$ 
         & $64(\pm21\%)$\\
c & $3.16(\pm7\%)$ & $1.23(\pm9\%)$ & $3.81(\pm6\%)$ & - 
  & $1.65(\pm7\%)$ & $4.3(\pm23\%)$  & $61(\pm29\%)$\\
\hline
\end{tabular}
 \parbox{4.8in}{
\begin{description}
\item{$*$}:
a: PS+Cyclohexane at $35^\circ$C. b: PS+CS$_2$ at $20^\circ$C.
 c: PS+Toluene at $25^\circ$C.
\end{description}
}
\end{center}
\end{table}
%
%
%
\subsection{PS+Cyclohexane at $35^{\circ}$C\label{resultpscyc}}
Eqn \ref{Omega.of.A} and \ref{Omegacyc}, together with the assumption that
 $\underline{P_\infty}=2$ yield
\begin{equation}
\{\Cmax A^{3-1/\nu}\}=1.95 \times 10^{-2}(\pm 7\%)\:\mbox{\AA$^{-2}$},
\label{cmaxacyc}
\end{equation}
where $\nu=0.5$.
Hence $\{A\} = 3.21(\pm 7\%)\mbox{\AA}$.
According to Eqn \ref{xi.phi.A.ratio},
\[
 \xi_\phi = \{A\}\cdot {\{\Cmax\} \over C} = 3.21\cdot{\{\Cmax\} \over C}(\pm 7\%)\:
\mbox{\AA}.
\]
Eqn \ref{xiscyc} can be rewritten as
\[
 \xi_s = 5.24 \cdot {\{\Cmax\} \over C} (\pm 5\%)\:\mbox{\AA}.
\]
Eqn  \ref{xidcyc} gives rise to
\[
 \xi_D = 22.5 \cdot{\{\Cmax\} \over C}(\pm 8\%)\:\mbox{\AA},\:\: 
  \mbox{for $C<4.6\times 10^{-4}$\AA$^{-3}$}.
\]
Eqn \ref{MRTPicyc} leads to
\[
 \xi_\Pi = 15.54 \cdot {\{\Cmax\} \over C}\:\mbox{\AA},
\]
It is now straightforward to calculate the $\xi$ ratios:
We find $\underline{\xi_\phi/\xi_s}=$
$0.61(\pm9\%)$, $\underline{\xi_D/\xi_s}=$ $4.29(\pm10\%)$, and
$\underline{\xi_\Pi/\xi_s}=$ $2.97(\pm5\%)$.
To calculate $\underline{\beta_2}$ and $\underline{\beta_4}$,
we combine Eqn \ref{S0Pi}, \ref{MRTPicyc}, \ref{cmaxacyc} and \ref{S001}, 
which yields $\underline{\beta_2}= 18(\pm 22\%)$.
Hence $\underline{\xi_\Pi / \xi_\phi} = 4.84(\pm8\%)$ in view of Eqn \ref{Piphi}.
Furthermore, from Eqn \ref{XIS}, we obtain
$\underline{\beta_4}= 968(\pm28\%)$.
 Eqn \ref{mom} then gives
\[
 \underline{r_n}\equiv \int_0^{\infty} dx\, x^n (\Czero{x\xi_\phi}/C-1)\goesto
 3(\pm22\%)\:\: \mbox{and}\:\: 48(\pm28\%)
\]
 for $n=$2 and 4, respectively.
%
%
%
\subsection{Perdeuterated PS+CS$_2$ at $20^{\circ}$C\label{SecPSCS2}}
Eqn \ref{Omega.of.A} and \ref{lamPSCS2} with $\underline{P_\infty}=1.23(\pm4\%)$
 give rise to 
$\{\Cmax A^{3-1/\nu}\}=2.34(\pm 8\%) \times 10^{-2}\: \mbox{\AA}^{-1/\nu}$ and
hence $\{A\}=2.82(\pm6\%) \mbox{\AA}$, whereas 
Eqn \ref{iqPSCS2} yields $\{A\}=2.75(\pm6\%)\mbox{\AA}$.
The average of these two $\{A\}$'s leads to
\[
 \xi_\phi = \{A\}\cdot \left({\{\Cmax\} \over C}\right)^{\frac{\nu}{3\nu-1}} 
= 2.785 \cdot \left({\{\Cmax\} \over C}\right)^{\frac{\nu}{3\nu-1}} (\pm 7\%)\:
\mbox{\AA},
\]
where $\nu=0.588$.
We also obtain $\underline{P_\infty}=1.20(\pm 11\%)$ using the same set of data
(Eqn \ref{lamPSCS2} and \ref{iqPSCS2}),
which agrees with the simulation results(Section \ref{sofchi}).
Several $\xi_s$'s for different concentrations are given in Section
 \ref{PSCS2}. 
From these data, we obtain $\underline{\xi_\phi / \xi_s}=1.23(\pm 8\%)$.
In addition, by virtue of Eqn \ref{Sxif}, we determine $\xi_f/\xi_s=1.27(\pm1\%)$,
which implies $\underline{\xi_f / \xi_\phi}=1.03(\pm 8\%)$.
Eqn \ref{XIS} and \ref{xif} then determine the reduced moments
$\underline{\beta_2}=4.8(\pm14\%)$ and $\underline{\beta_4}=64(\pm21\%)$. 
Alternatively, they determine the $\underline{r_n}$ moments
defined in Eqn \ref{mom}:
$\underline{r_2}=1.2(\pm14\%)$ and $\underline{r_4}=4.5(\pm21\%)$.
%
%
\subsection{PS+Toluene at $25^{\circ}$C \label{PStoluene}}
Eqn \ref{Omega.of.A} and \ref{Omegatol} give 
$\{\Cmax A^{3-1/\nu}\}=0.0273(\pm9\%) \mbox{\AA$^{-1/\nu}$}$, 
where we set $\nu=0.59$ and $\underline{P_\infty}=1.23(\pm4\%)$.
This in turn yields $\{A\}=3.16(\pm7\%)\mbox{\AA}$ and
\[
 \xi_\phi = \{A\}\cdot \left({\{\Cmax\} \over C}\right)^{\frac{\nu}{3\nu-1}} 
= 3.16 \cdot \left({\{\Cmax\} \over C}\right)^{0.77} (\pm 7\%)\:
\mbox{\AA}.
\]
Eqn \ref{xisrhotol} can be written as
\[
 \xi_s = 2.57\cdot\left({\{\Cmax\} \over C}\right)^{0.77}(\pm 5\%)\:\mbox{\AA}.
\]
Eqn \ref{Sxpirhotol} gives
\[
 \xi_\Pi = 9.80\cdot\left({\{\Cmax\} \over C}\right)^{0.77}(\pm 3\%)\:\mbox{\AA}.
\]
$\xi_D$ is determined from Eqn \ref{dcc}:
\[
 \xi_D = 4.25 \cdot\left({\{\Cmax\} \over C}\right)^{0.75}(\pm 5\%)\:\mbox{\AA},
\:\: \mbox{for $0.02 \leq C/ \{\Cmax\} \leq 0.08$}.
\]
The above results give $\underline{\xi_\phi / \xi_s}=1.23(\pm9\%)$,
 $\underline{\xi_D/\xi_s}=1.65(\pm7\%)$, and
 $\underline{\xi_\Pi/\xi_s}=3.81(\pm6\%)$.
By identifying Eqn \ref{Sxpirhotol} with Eqn \ref{S001}, 
we obtain $\underline{\beta_2}= 4.3(\pm 23\%)$, which yields 
$\underline{\xi_\Pi/\xi_\phi}=3.16(\pm8\%)$(Eqn \ref{Piphi}).
Eqn \ref{XIS} then gives $\underline{\beta_4}= 61(\pm29\%)$.
These results are consistent with those of PS+CS$_2$ presented in the previous section.
Furthermore, Eqn \ref{mom} gives rise to
$\underline{r_2}=1.1(\pm23\%)$ and $\underline{r_4} = 4.3(\pm29\%)$.
%
%
%
\section{Discussion\label{Discussion}}
\subsection{Solvent Quality}
In good-solvent cases such as PS+CS$_2$ at $20^{\circ}$C,
the relations between $\xi_s$, $\xi_\phi$, $\xi_f$, and $\xi_\Pi$(Eqn
\ref{XIS}, \ref{Piphi} and \ref{xif}) become strictly correct only 
for a sufficiently good solvents.
Whether the solvent is good enough may be expressed 
using the notion of the thermal blob\cite{Dao}.
 The size of the thermal blob $\{\xi_T\}$
marks the crossover between ideal chain and 
self-avoiding behavior. 
 For chain sections of size much smaller than 
$\{\xi_T\}$ the polymer behaves as an ideal chain; for sections much larger
 than $\{\xi_T\}$, 
it exhibits excluded-volume expansion.  
In such solvents, the good-solvent scaling 
properties disappear for high concentrations such that $\xi_\phi \leq 
\{\xi_T\}$.  However, the good-solvent behavior appears for lower concentrations such that 
$R_G \gg \xi_\phi \gg \{\xi_T\}$.  Accordingly, the universal ratios
reported in rows b and c of Table \ref{tabratio} should hold only in this same regime.  
To be more specific, when $R_G \gg \xi_\phi \gg \{\xi_T\}$, 
the first-order correction to those $\xi$ ratios   
 goes as some positive power of $\{\xi_T\}/\xi_\phi$.
If $\xi_\phi$(or $R_G$) is of the order of $\{\xi_T\}$, \ie, the system is in
the marginal-solvent condition, special care must be taken to 
interpret experimental data\cite{Han}.  
%
%
\subsection{Consistency with Previously Reported Universal Ratios}
In this work, we demonstrate the universal properties of semidilute polymer
solutions by showing the constancy of those $\xi$ ratios without referring to
the dilute properties such as $R_G$. However, there exists an equivalent alternative,
which relies on the scaling relations between dilute and semidilute properties.

In the appendix,
we obtain a scaling formula for osmotic pressure $\Pi$(Eqn \ref{NkTPi}) and
define a universal constant $\underline{k_\Pi}$, 
which is related to the reduced moment $\underline{\beta_2}$(Eqn \ref{kPi}).
Since $\underline{P_\infty}$ and $c_{_0}$ are known, and the reduced moment
$\underline{\beta_2}$ is given in Table \ref{tabratio},
we determine $\underline{k_\Pi}=2.2(\pm22\%)$ and $9.4(\pm24\%)$ for theta
and good solvents, respectively.
For good-solvent cases where $\nu=0.588$, 
des Cloizeaux and Noda obtained $\underline{k_\Pi} \approx 9.85$ using
the renormalization group theory(to first order in $\epsilon$)\cite{Clo}.
Noda et al reported $\underline{k_\Pi}=10.0$ and $\nu=0.585$ for
 poly($\alpha$-methylstyrene) in toluene\cite{Noda}.
Adam et al determined $\underline{k_\Pi}=9.7$ and $\nu=0.586\pm 0.006$ for 
semidilute polyisoprene in cyclohexane\cite{MAdam}.
Their results together with ours confirm the universality of $\underline{k_\Pi}$,
and hence that of the ratio $\underline{\xi_\Pi/\xi_\phi}$ for good-solvent cases(Eqn \ref{ratio.xipi.xiphi}).

Des Cloizeaux and Jannik defined the Kuhnian overlap length $\xi_k$ as\cite{PiS}
\[
  \xi_k = X (CX^3/N)^{\nu/(1-3\nu)},
\]
where $X^2=R_e^2/3=2R_G^2/\underline{\aleph}$ in the dilute limit.
 $R_e$ is the end-to-end distance of a dilute polymer.
The numerical constant $\underline{\aleph} \equiv 6R_G^2/R_e^2\simeq 0.952$
for an isolated polymer in a good solvent\cite{PiS}.
$\xi_k$ can be rewritten in a form similar to Eqn \ref{xiPiRG}:
\[
  {\xi_k \over R_G} = \left(\underline{\aleph}/ 2\right)^{1\over{2(3\nu-1)}}
    \left(\frac{C^*}{C}\right)^{{\nu}\over{3\nu-1}},
\]
where $C^*\equiv N/R_G^3$ is the overlap concentration.
Dividing  Eqn \ref{xiphiRG} by the above formula gives rise to
\[
 \underline{\xi_\phi/\xi_k} = 
  \left( \underline{P_\infty}/c_{_0} \right)^{{\nu}\over{3\nu-1}}
   \left(\underline{\aleph}/ 2\right)^{-1\over{2(3\nu-1)}}.
\]
This implies $\underline{\xi_\phi/\xi_k} = 0.318$ and $0.243$ for theta and good 
solvents respectively.
In addition, by use of $\underline{\xi_\phi/\xi_s}$ given in Table \ref{tabratio},
we determine $\underline{\xi_s/\xi_k} = 0.522(\pm 9\%)$ for theta cases, 
and $0.198(\pm 9\%)$ for good-solvent cases.
 The latter value agrees with that($0.18$) reported in their
 book(Chapeter 15 of \cite{PiS}).
%
%
\subsection{Dynamic Length $\xi_D$\label{dis.xiD}}
While the universality of the static ratios $\underline{\xi_\phi / \xi_s}$ and
 $\underline{\xi_\Pi / \xi_s}$ are robust, that of the dynamic ratio
  $\underline{\xi_D / \xi_s}$
is not. For semidilute polymer solutions, $\xi_D$  goes as
 $C^{-\alpha}$, where the theoretical value of $\alpha=0.77$ for
good-solvent conditions. 
 Experiments, however,  often report $\alpha=0.5-0.75$\cite{Roo,Bro,Wil}. 
The deviation of the exponent  from the scaling theory prediction
suggests the underlying dynamics of the co-operative diffusion of 
semidilute polymers
might be more complex than the scaling theory can explain.
Several attempts have been made to resolve this discrepancy
by attributing it to the cross-over behavior of
 the system under investigation\cite{Han,Wil,Shiwa}.
Despite of these controversies,
we expect the scaling theory prediction should still hold in the 
semidilute limit, where $C^*\ll C \ll \{\Cmax\}$ and $N\goesto \infty$,
and this is confirmed by the experimental data collected in 
Schaefer and Han's review paper\cite{Han}. 
In this paper, we use their data to calculate  $\underline{\xi_D / \xi_s}$
 for PS+Toluene(Eqn \ref{dcc}).
%
\section{Conclusions}
Table \ref{tabratio} summarizes the ratios of those $\xi$'s
for semidilute polymers under theta and good-solvent conditions.
These ratios are expected to be universal in the semidilute regime where
$C^*\ll C \ll \{\Cmax\}$ and the chainlength $N\goesto \infty$.
The static lengths $\xi_s$, $\xi_\phi$, $\xi_f$ and $\xi_\Pi$ are 
related to each other through the reduced moments $\underline{\beta_2}$ and
 $\underline{\beta_4}$ 
 of the local concentration $\Czero{r}$(Section \ref{betas}).
The universality of these  length ratios is just the manifestation
of the scaling property of $\Czero{r}$.
The data in Table 2 provide a satisfying confirmation of this universality.  
Each of the reported ratios (except those involving $\xi_D$) was 
obtained from data over a substantial range of concentration---approaching an order of magnitude. 
 The ratios were independent of concentration to a degree given by the quoted uncertainty. 
 Further, the reduced moments $\underline{\beta_2}$ and $\underline{\beta_4}$ were
 independently obtained for two solvents, CS$_2$ and toluene.  The values obtained are very 
similar for the two solvents, further supporting the claim of universality.
As mentioned in Section \ref{intro}, these ratios are valuable since 
they allow us to infer one quantity such as $\{A\}$ or $\xi_\phi$ 
from another, such as the osmotic pressure $\Pi$.  
Moreover, $\underline{\beta_2}$ and $\underline{\beta_4}$ provide the 
information about the cross-over
behavior of $\Czero{r}$.
It might be a challenge for polymer theorists who are interested in 
$\Czero{r}$ to work out $\underline{\beta_n}$(or $\underline{r_n}$) 
and determine these ratios. 

As noted in Section \ref{dis.xiD},
the concentration dependence of $\xi_D$ may deviate from the scaling theory
prediction.
It seems to us that, so far there is no satisfactory explanation for this
discrepancy.
Thus, further theoretical and experimental investigations 
are needed.
%
%
%
%
\clearpage
\appendix
\noindent{\Large \bf Appendix}\vspace{0.3cm}\\
%
%
%
%
%
\noindent{Eqn \ref{sofq.czero} can be rewritten as}
\[
 S(q) = \int_{V} d^{3}r (\Czero{r} -C)\exp (i\vec{q}\cdot\vec{r})
\]
since the Fourier transform of any constant is zero as long as $q\neq 0$.
$S(q)$ can then be Taylor-expanded with respect to $q$(\cf Eqn \ref{TEs}):
\[
  S(q)=S_{0} \left( 1-\frac{1}{2} \frac{ \int d^{3}r \left(
\vec{q}\cdot\vec{r}\right)^2 \left( \Czero{r} - C \right) }{\int d^{3}r 
\left( \Czero{r} -C \right)} \cdots \right),
\]
where $S_0 = \int d^{3}r \left( \Czero{r} -C \right)$.
A straightforward calculation leads to
\begin{equation}
\xi_{s}^{2} = \frac{1}{6} \frac{ \int_{0}^{\infty} dr \, r^4 
\left( \Czero{r} -C \right) }{\int_0^{\infty} dr \, r^2\left( \Czero{r} -C \right)}.
     \label{xis2}
\end{equation}
By introducing $\underline{\beta_n}$(Eqn \ref{def.beta.n}) and using Eqn \ref{xi.phi.A.ratio},
\begin{eqnarray}
 S_0 &\goesto& 4\pi \underline{\beta_2}  \integral_{C_A > C} dr\, r^2 (C_A(r)-C)  \nonumber \\
        &=& \frac{4\pi}{3}(3\nu-1)\underline{\beta_2}
                            \{\Cmax A^{3-1/\nu}\}\xi_{\phi}^{1/\nu} 
      \label{S00}  \\
        &=& \frac{4\pi}{3}(3\nu-1)\underline{\beta_2} 
                   \{\Cmax A^{3-1/\nu}\}^{\frac{3\nu}{3\nu-1}} C^{\frac{-1}{3\nu-1}}
      \label{S001}
\end{eqnarray}
Incorporating Eqn \ref{Siq0}, \ref{Pinfty} and
the overlap concentration $C^*\definedas N/R_G^3$ into
Eqn \ref{S001} leads to
\[
 S_0 \goesto  \frac{4\pi}{3} (3\nu-1) 
\left( \underline{P_\infty}/c_{_0} \right)^{{3\nu}\over{3\nu-1}} 
 \underline{\beta_2}  \left(C/C^*\right)^{\frac{1}{1-3\nu}} N,
\]
where $c_{_0}$ is defined below Eqn \ref{Siq0}.
Using this $S_0$ and Eqn \ref{S0Pi}, we obtain
\begin{equation}
   \frac{N}{k_B T} \frac{\Pi}{C} \goesto \underline{k_\Pi} \left(
\frac{C}{C^*}\right)^{\frac{1}{3\nu-1}},  \label{NkTPi}
\end{equation}
where
\begin{equation}
 \underline{k_\Pi} =(4\pi\nu \underline{\beta_2})^{-1} 
 \left( \underline{P_\infty}/c_{_0} \right)^{{3\nu}\over{1-3\nu}}.
  \label{kPi}
\end{equation}
Furthermore, Eqn \ref{NkTPi} gives
\begin{equation}
\frac{\xi_\Pi}{R_G} \goesto  \underline{k_\Pi}^{-1/3}
\left(\frac{C^*}{C}\right)^{{\nu}\over{3\nu-1}} \label{xiPiRG}
\end{equation}
Similarly, Eqn \ref{xi.phi.A.ratio} can be expressed as
\begin{equation}
\frac{\xi_\phi}{R_G} \goesto \left( \underline{P_\infty}/c_{_0} \right)^{{\nu}\over{3\nu-1}}
\left(\frac{C^*}{C}\right)^{{\nu}\over{3\nu-1}}   \label{xiphiRG}
\end{equation}
Combining Eqn \ref{xiPiRG} with \ref{xiphiRG} yields
\begin{equation}
 \underline{\xi_\Pi/\xi_\phi} \goesto 
\underline{k_\Pi}^{-1/3}\left( \underline{P_\infty}/c_{_0} \right)^{{\nu}\over{1-3\nu}} 
 =(4\pi\nu \underline{\beta_2})^{1/3}. \label{ratio.xipi.xiphi}
\end{equation}
By the same token, Eqn \ref{xis2} is reduced to
\begin{equation}
    {\xi_{s}}^{2} \goesto {\xi_{\phi}}^{2} \left[ \frac{1}{6}
\frac{\left( \frac{\nu}{1+2\nu}-\frac{1}{5}\right)\underline{\beta_4}}{\left(
\nu-\frac{1}{3}\right)\underline{\beta_2}}\right].
    \label{sphi}
\end{equation}
We may define another kind of 
dimensionless moment $\underline{r_n}$ for $\Czero{r}$, which carries
the same information as does $\underline{\beta_n}$: 
\begin{equation}
\underline{r_n} \equiv 
\int_0^{\infty} dx\, x^n
\left(\frac{\Czero{x\xi_\phi}}{C}-1\right)\goesto
  \frac{(3-1/\nu)\underline{\beta_n}}{(n+1)(n-2+1/\nu)}, \label{mom}
\end{equation}
where $x=r/\xi_{\phi}$.
%
%
%
%

%
%
%
\end{document}